# Latent Topic Models for Hypertext

**Amit Gruber**
School of CS and Eng.
The Hebrew University
Jerusalem 91904 Israel
amitg@cs.huji.ac.il

**Michal Rosen-Zvi**
IBM Research Laboratory in Haifa
Haifa University, Mount Carmel
Haifa 31905 Israel
rosen@il.ibm.com

**Yair Weiss**
School of CS and Eng.
The Hebrew University
Jerusalem 91904 Israel
yweiss@cs.huji.ac.il

## Abstract

Latent topic models have been successfully applied as an unsupervised topic discovery technique in large document collections. With the proliferation of hypertext document collection such as the Internet, there has also been great interest in extending these approaches to hypertext [6, 9]. These approaches typically model links in an analogous fashion to how they model words - the document-link co-occurrence matrix is modeled in the same way that the document-word co-occurrence matrix is modeled in standard topic models.

In this paper we present a probabilistic generative model for hypertext document collections that explicitly models the generation of links. Specifically, links from a word $w$ to a document $d$ depend directly on how frequent the topic of $w$ is in $d$, in addition to the in-degree of $d$. We show how to perform EM learning on this model efficiently. By not modeling links as analogous to words, we end up using far fewer free parameters and obtain better link prediction results.

## 1 Introduction

The need to automatically infer the different topics discussed in a corpus arises in many applications ranging from search engines to summarization software. A prominent approach is modeling the corpus with a latent topic model where each document is viewed as a mixture of latent topics or factors, and the factors, shared by the whole corpus, are related to the terms or words appearing in the documents.

Many of the topic models share the "bag of words" assumption where each document is represented as a histogram of terms, ignoring the order of terms and the internal structure of the documents. The entire corpus is represented as a document-term co-occurrence matrix. Semantic analysis is done by projecting the document-term co-occurrence matrix onto a lower dimensional factor space. In algebraic methods such as Latent Semantic Analysis [7] it is projected onto a linear factor space using SVD. In statistical methods such as Probabilistic LSA [13], Latent Dirichlet Allocation [3] or the somewhat more general formalism, Discrete PCA [4] the document-term co-occurrence matrix is projected onto a simplex by maximizing the observations likelihood. In recent years these latent topic models have been extended in various ways. In particular, correlation between topics [2] and their dynamics over time [1] have been directly modeled. The use of additional information provided in the corpus such as authorship information has been studied [18]. In addition, novel models that depart from the bag of words assumption and do consider the internal ordering of the words in sentences within a document have been developed. These models combine local dependencies in various ways; for example, combining n-grams with a hierarchical topic model [19], modeling syntax [10] and modeling the continuous drift from one topic to another within a document [12]

In this paper, we address the question of how to enrich the model by considering links between documents, such as hyperlinks in hypertext or citations in scientific papers. With the emergence and rapid growth of the World Wide Web, hypertext documents containing links to other documents have become ubiquitous. The connectivity between documents has proven to play an important role in determining the importance and relevance of a document for information retrieval or the interest of a certain user in it [17, 5, 14]. In particular, Dietz at al. [8] have recently proposed a generative topic model for the prediction of citation influences, called the citation influence model. It models the particular structure of paper citations where the citations graph can be described by a directed acyclic graph

(DAG); a setting that does not hold in the case of the World Wide Web and other hypertext corpora.

There are few previous works that extend topic models to include link information. Cohn and Hofmann [6] introduce a joint probabilistic model for content and connectivity. The model is based on the assumption that similar decomposition of the document term co-occurrence matrix can be applied to the cite-document co-occurrence matrix in which each entry is a count of appearances of a linked-document (or citation) in a source document. In this approach, links are viewed as additional observations and are analogous to additional words in the vocabulary, but with different weight when estimating the topic mixture of the document. Erosheva et al. [9] also makes use of a decomposition of term-document and citation-document co-occurrence matrices by extending the LDA model to include a generative step for citations. Note that these models only learn from the co-occurrence matrix of citations without exploiting the information conveyed by the cited documents text. Thus, if the citation-document co-occurrences matrix is very sparse, the generalization power of the models is very limited.

In this paper, we suggest a novel generative model for hypertext document collection that we name the latent topic hypertext model (LTHM). Our approach includes direct modeling of real-world complex hypertext collections in which links from every document to every document may exist, including a self-reference (a document linking to itself). We model a link as an entity originating from a specific word (or collection of words) and pointing to a certain document. The probability to generate a link from a source document $d$ to a target document $d'$ depends on the topic of the word from which the link is originating, on the importance of the target document $d'$ (estimated roughly by the in-degree) and on the topic mixture of the target document $d'$. In this way, an observed link directly affects the topic mixture estimation in the target document as well as the source document. Moreover, the non-existence of a link between two documents is an observation that serves as evidence for the difference between the topic mixtures of the documents.

We introduce the LTHM and related models in Section 2 and describe the approximate inference algorithm in Section 3. Experimental results obtained by learning two datasets are provided in Section 4. Finally, we discuss the results in Section 5.

## 2 The latent topic hypertext model

The topology of the World Wide Web is complicated and unknown. The corpus we work with is a subset of the World Wide Web and its topology can be arbitrary accordingly. By no means can we assume it forms a DAG. Therefore, we would like to allow each document to link to any other document, allowing for loops, i.e. directed cycles of links originating in a certain document and ending in the same document. In particular, we would like to allow for self loops with links where a document links to itself. The solution is a generative model that consists of two stages. In the first stage, the document content (the words) is created. After the text of all the documents has been created, the second stage of creating links takes place.

The contribution of this paper is in modeling link generation and suggesting an approximate inference algorithm for studying it. The text in the documents can be generated using several of the various models mentioned in section 1. For simplicity, we describe text generation (and inference, accordingly) using LDA [3]. In the following section, we first briefly review the LDA model (2.1). Second, we describe the second stage of link generation (2.2). Finally, we discuss related models (in section 2.3).

### 2.1 Document generation (LDA)

According to the LDA model, a collection of documents is generated from a set of $K$ latent factors or topics. One of the main assumptions in the model is that for each topic there is a single multinomial random variable $\beta$ that defines the probability for a word given a topic for all documents in the collection. Each document is characterized by a particular mixture of topic distribution defined by the random variable $\theta$. The generation of the $N_d$ words of each document $d$ in a corpus contains two stages: first, a hidden topic $z$ is selected from a multinomial distribution defined by $\theta$. Second, given the topic $z$, a word $w$ is drawn from the multinomial distribution with parameters $\beta_z$. Figure 2.1a illustrates the generative model.

Formally, the model can be described as:

1. For each topic $z = 1, ..., K$ choose $W$ dimensional $\beta_z \sim \text{Dirichlet}(\eta)$

2. For each document $d = 1, ..., D$

   Choose $K$ dimensional $\theta \sim \text{Dirichlet}(\alpha)$

   For each word $w_i$, indexed by $i = 1, ..N_d$

   Choose a topic $z_i^W \sim \text{Multinomial}(\theta_d)$

   Choose a word $w_i \sim \text{Multinomial}(\beta_{z_i^W})$

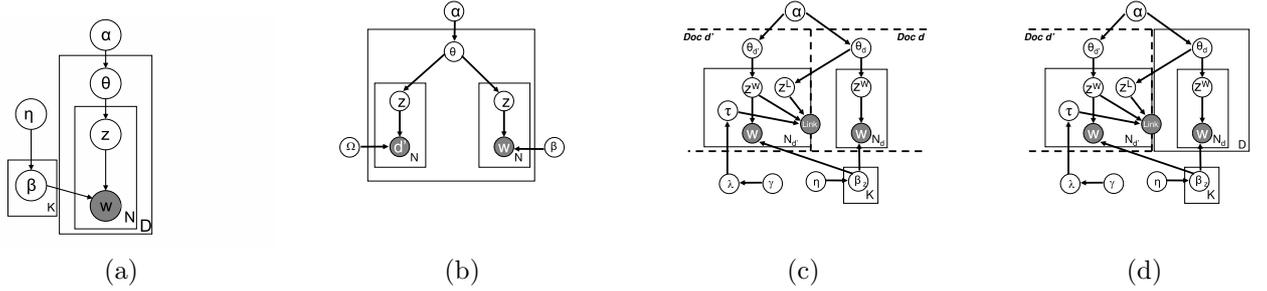

Figure 1: **a.** The LDA model. **b** The link-LDA model. **c.** The LTHM model in a scenario of generating links from document $d'$ to document $d$. **d.** The LTHM model in a scenario of generating links from document $d'$ to any other document in the collection of $D$ documents.

### 2.2 Link generation

We assume that links originate from a word, and each word can have at most one link associated with it[1]. For simplicity, we restrict the discussion to the case where a link is anchored to a single word. The generalization to the case where the link is anchored to a sequence of words can be carried out by forcing the topics of these words to be identical, as proposed in [12]. The generation of links is carried out by iterating over all the words in the document and for each word determining whether to create a link and if so, what is the target document.

Let us limit the discussion first to the case where the corpus contains two documents, $d$ and $d'$, and links are generated from words in document $d'$ to document $d$. When iterating over the words in $d'$, at the $i^{th}$ word, we need to decide whether to create a link from $w_i$ to $d$ or not. This decision contains two steps (at most, as sometimes the first step is sufficient to determine that no link needs to be created). The first step is drawing at random a variable $\tau_i$ from a multinomial $\lambda$. In general, $\tau_i$ can take values from 0 to $D$, and in this degenerated example it can take two values: 0 indicates no link and $d$ indicates a link to document $d$. Only if $\tau_i = d$ do we consider adding a link to document $d$ and then proceed to the next step, which is randomly drawing the topic of the link, $z^L$. The topic assignment $z^L$ is drawn from $\theta_d$, the mixture of topics of the document $d$. A link is created iff $z_i^W = z^L$, Figure 2.1c illustrates the full generative model for this degenerated example.

The generalization to the (still degenerate) case of generating links from a single document $d'$ to any other document in a collection of $D$ documents is illustrated in Figure 2.1d. In this case, the generation of links from words in document $d'$ starts by select-ing $\tau_i \in \{1 \ldots D, \emptyset\}$ for every word $i = 1..N_{d'}$ of the document $d'$. $\tau$ is drawn at random from $\lambda \in R^{D+1}$, a multinomial distribution indicating the probability of considering a link to each one of the $D$ documents or not having a link at all. It is a measure of the importance (or in-degree) of the documents in the corpus. $\lambda$ itself is drawn from the hyperparameter $\gamma$. The Dirichlet prior $\gamma \in R^{D+1}$ is not symmetric and favors not creating links, as most words do not have an associated link: $\gamma_i \ll \gamma_\emptyset$ for $i = 1 \ldots D$. Also, note that links from a document to itself are allowed in this model (as well as in real life).

---
[1]If a link is anchored to an image, for example, we could substitute a fictitious word for that link.

The most general case, in which every document can contain words linked to any other document, is generated by sequentially going through all words and all documents and drawing at random the corresponding $\tau$s and $z^L$s in the way described above.

Formally, the generative process is:

1. Choose $D + 1$ dimensional $\lambda \sim \text{Dirichlet}(\gamma)$
2. For each document $d = 1, ..., D$

    For each word $w_i$, indexed by $i = 1, ..N_d$

    Choose $\tau_i \in \{1 \ldots D, \emptyset\} \sim \text{Multinomial}(\lambda)$

    If $\tau_i \neq \emptyset$ choose a topic $z_L \sim \text{Multinomial}(\theta_{\tau_i})$

    If $z^L = z_i^W$ create a link $L_i = \tau_i$ from

    word $i$ to document $\tau_i$

### 2.3 Related Models

Both models of [6] and [9] – that we refer to as link-PLSA and link-LDA, respectively, following [16]'s suggestion – are make of the citation-document co-occurrence matrix in a similar manner. We focus on the link-LDA model that is somewhat closer to our model. According to this approach two types of observed variables are modeled: words in documents and citation in documents. The generation of these variables is carried out by first selecting a mixture of topics for each of the documents and then for each of the words and citations on the document generating a hidden topic from which the observation is selected at random from the $\beta_z$ in the case of words and from $\Omega_z$ in the case of citations; Here $z = 1, ..., K$. The model is illustrated in Figure 2.1b.

Formally, the model can be described as:

For each document $d = 1, ..., D$
    Choose $K$ dimensional $\theta \sim \text{Dirichlet}(\alpha)$
    For each word $w_i$, indexed by $i = 1, ..N_d$
        Choose a topic $z_i \sim \text{Multinomial}(\theta_d)$
        Choose a word $w_i \sim \text{Multinomial}(\beta_{z_i})$
    For each citation $d'_i$, indexed by $i = 1, ..L_d$
        Choose a topic $z_i \sim \text{Multinomial}(\theta_d)$
        Choose a citation $d'_i \sim \text{Multinomial}(\Omega_{z_i})$

Note that in the LTHM, the probability to create a link given topic is $\Pr(link = d|z) = \lambda_d \theta_d(z^W)$. There are only $D$ additional parameters $\lambda_1 \ldots \lambda_D$ to denote the document importance for link creation, whereas in the link-LDA model there are $DK$ additional parameters, $\Omega_{d,z} = \Pr(d' = d|z)$. Also, in LTHM, the very existence or non-existence of a link is an observation, while this is not explicitly modeled by the link-LDA. Moreover, according to the LTHM, a link shares the same topic with the word it originates from and at the same time affects the topic mixture in the cited document.

## 3 Approximate Inference

Exact inference in hierarchical models such as LDA and PLSA is intractable due to the coupling of the latent topics and the mixing vectors $\beta, \theta$. The hypertext model presented in this paper shares this coupling and adds a unique coupling between topic mixing vectors; hence, exact inference is intractable in it as well. In recent years, several alternatives for approximate inference in such models have been suggested: EM [13] or variational EM [3], Expectation propagation (EP) [15] and Monte-Carlo sampling [18, 11]. Unlike other hypertext topic models, in LTHM not only the identities of the ends of a link are observations, but also the link's very existence (or non-existence). Taking into account the non-existence of links in sampling-based inference necessitates further approximations. We therefore perform inference using EM.

EM deviates from fully Bayesian methods by distinguishing between latent variables and parameters of the model. The latent variables are the latent topic of a word, $z^W$, the latent topic involved in link generation, $z^L$, and the variable $\tau$. The parameters of the model are the topic mixing vectors $\theta_d$, the word mixing vectors $\beta_z$ and the document link importance parameter $\lambda_d$. The Dirichlet hyperparameters $\alpha$, $\eta$ and $\gamma$ are fixed.

In the link generation process, unless a link is created, the value of $\tau$ is unknown. It might have not been created because $\tau = \emptyset$ or because of topic mismatch between the source document and any other document. For this reason, we need to consider all possible options with their probability during inference: for each source document $d$ and each word in it from which there is no outgoing link, we need to consider all $D$ possible $z^L$ variables. The number of the potential latent variables $z^L$ is $D \sum_d Nd$ which is quadratic in the number of documents. It is therefore infeasible to compute explicitly the posterior distribution of each one of these latent variables. However, in the M-step, only aggregations of these posterior distributions are needed. The required aggregations can be computed efficiently (in time linear in the size of the corpus) by taking advantage of symmetries in the model as described in section 3.2 and in the appendix. We begin with the M-step equations, detailing what are the required expectations. Then we describe how the required posteriors and aggregations are computed in the E-step.

### 3.1 M-step

In the M-step, MAP estimators for the parameters of the model, $\theta_d, \beta_z$ and $\lambda$ are found. Let $G_{z,w}$ denote the number of occurrences of a word $w$ with topic $z^W = z$. The update rule for $\beta_{z,w}$ is identical to that in standard LDA:

$$\beta_{z,w} \propto E(G_{z,w}) + \eta_w - 1 \qquad (1)$$

The MAP estimator for $\theta_d$ takes into account topics of words and links that were drawn from $\theta_d$. The word topics, $z^W$, are drawn from $\theta_d$ for each of the words in document $d$. The link topics are the topics $z^L_{d',i,d}$ drawn from $\theta_d$ when considering a link from any other document $d'$ to $d$. These are the cases where $\tau_{d',i} = d$ for any $d', i$ regardless of whether the link has been created or not. For the purpose of inference, we count the topics $z^L_{d',i,d}$ separately for links and for non-links. Let $F_{d,z}$ denote the number of occurrences of a topic $z$ associated with any word in document $d$. Let $V_{d,z}$ be the number of occurrences of a topic $z$ associated with any incoming link of document $d$. Let $U_{d,z}$ be the number of times $\tau_{d',i} = d$ but the topic generated for the link by document $d$, $z^L_{d',i,d}$, does not match the topic of the $i$th word in the the document $d'$, $z^W_{d',i}$ and therefore a link has not been created.

$$\theta_{d,z} \propto E(F_{d,z}) + E(V_{d,z}) + E(U_{d,z}) + \alpha_z - 1 \quad (2)$$

Note that in the standard LDA model, we would have just the first term (the expected number of times topic z appears in document d) and the Dirichlet prior. In the LTHM, we add two more terms which model the influence of links (or non-links) on the topic distribution.

The computation of $E(V_{d,z})$ and $E(U_{d,z})$ is described in section 3.2.

The MAP estimator for $\lambda$ is

$$\lambda_d \propto E(T_d) + \gamma_d - 1 \quad (3)$$
$$\lambda_\emptyset \propto \sum_d N_d - \sum_d E(T_d) + \gamma_\emptyset - 1 \quad (4)$$

Where $T_d$ is the number of times that $\tau_{d',i} = d$ for any $d'$ and any word $i$ in it (this includes the case of $d' = d$ where a self link is considered). Notice that $T_d = \sum_z (V_{d,z} + U_{d,z})$. The normalization factor in equations 3 and 4 includes the term $\lambda_\emptyset$, the most frequent case that there is no link at all.

### 3.2 E-step

In the E-step, expectations required for the M-step are computed with respect to the posterior distribution of the latent variables. The expectations required for the M-step are $E(G_{d,z}), E(F_{z,w}), E(V_{d,z}), E(U_{d,z})$ and $E(T_d)$.

$E(G_{d,z})$ is the expected number of occurrences of a topic $z$ in document $d$ as a topic of word and $E(F_{z,w})$ is the expected number of occurrences of a word $w$ with topic $z$:

$$E(G_{d,z}) = \sum_{i=1}^{N_d} \Pr(z^W_{d,i} = z | \bar{w}, \bar{L}) \quad (5)$$

$$E(F_{k,z}) = \sum_{d=1}^{D} \sum_{i=1}^{N_d} \Pr(z^W_{d,i} = z, w_{d,i} = w | \bar{w}, \bar{L}) \quad (6)$$

where $\bar{w} = w_1 \ldots w_{N_d}$ and $\bar{L} = L_1 \ldots L_{N_d}$. The posterior distribution of $z^W$ is explicitly computed, taking into account words and links (or the non-existence of a link) as observations.

$$\Pr(z^W_{d,i} = z | \bar{w}, \bar{L}) \propto \theta_d(z) \Pr(L_{d,i} | z^W_{d,i} = z) \phi_z(w_{d,i}) \quad (7)$$

where $\Pr(L_{d,i} | z^W_{d,i} = z)$, the probability of a link observation is

$$\Pr(link(d,i) \to d' | z^W_{d,i} = z; P) = \lambda_{d'} \theta_{d'}(z) \propto \theta_{d'}(z)$$

if a there is a link from word $i$ in document $d$ to document $d'$, and

$$\Pr(no - link(d,i) | z^W_{d,i} = z; P) = 1 - \sum_{d'} \lambda_{d'} \theta_{d'}(z)$$

if there is no link associated with word $i$ in document $d$.

Naïve computation of $E(V_{d,z})$ and $(E(U_{d,z})$ would require estimating the posterior distributions of $\tau_{d',i}$ and $z^L_{d',i,d}$ for all triplets $(d',i,d)$. As mentioned before, explicit computation of these posteriors is infeasible due to large number of these variables. Rather than computing this posterior distribution explicitly over $z^L$ and $\tau$, only the aggregations $E(V_{d,z}), E(U_{d,z})$ are computed.

$E(V_{d,z})$ is the expected number of occurrences of links incoming to document $d$ with topic $z$. In the case where a link exists, the posterior distributions of $z^L$ and the corresponding $z^W$ are equal; hence, $V_{d,z}$ can be computed by summing posterior probabilities of $z^W$:

$$E(V_{d,z}) = \sum_{(d',i) \in A_d} \Pr(z^L_{d',i,d} = z | O, P) \quad (8)$$
$$= \sum_{(d',i) \in A_d} \Pr(z^W_{d',i} = z | O, P)$$

where $A_d = \{(d',i) : link(d',i) \to d\}$, $O$ is the set of all observations and $P$ is the model parameters.

$E(U_{d,z})$ is the expected number of times $\tau_{d',i} = d$ for any $d',i$ in the corpus, but $z^L_{d',i,d} \neq z^W_{d',i}$. The basic idea in the computation of $E(U_{d,z})$ is that it factors into topic dependent terms and document-topic dependent terms. The topic dependent terms can be computed in a single linear pass over the corpus (in each iteration). The document-topic dependent terms are specific to each $U_{d,z}$. Combining them with the topic dependent terms to compute $U_{d,z}$ is done in a constant number of operations (for each $d, z$). The computation is detailed in the appendix.

Finally, after $E(V_{d,z})$ and $E(U_{d,z})$ have been computed,

$$E(T_d) = \sum_z [E(V_{d,z}) + E(U_{d,z})] \quad (9)$$

Despite the quadratic number of latent variables, the runtime of both E and M steps is linear in the size of the corpus times the number of topics.

There are a number of extensions to the LDA model that can be considered here, as the approximate inference algorithm described above can be easily adapted for many of the alternatives mentioned in section 1. For example, suppose one wishes to model the text with the HTMM [12], the difference would be in the computation of the posterior of word topics, $\Pr(z^W|w_1 \ldots w_{N_d}, L_1 \ldots L_{N_d})$. In HTMM, this posterior would be computed using the forward-backward algorithm, considering both words and links as the topic emission probabilities. Alternatively, if one wishes to make use of authorship information by applying the Author-Topic model [18], it would require to consider an additional latent variable $x$ for the authorship of each word and compute posterior probabilities $\Pr(x, z^W|w_1 \ldots w_{N_d}, L_1 \ldots L_{N_d})$ and modify the definition of $\theta_d$. Yet, the modeling of links stays very similar; in addition to latent topic of the link, $z^L$, only a latent author to the link needs to be selected, $x_L$.

## 4  Experiments

In this section, we explore the relations between links and topics discovered by LTHM and evaluate its predictive power with respect to links. We compare LTHM's link prediction with previous approaches for combining links: link-PLSA[6] and link-LDA[9]. We also compare to link prediction by a non-topic method, based only on the frequency a web page is linked, ignoring the contents of the text in the corpus. For this comparison, we use two datasets of web pages: the webkb dataset (8282 documents with 12911 links) and a small dataset of Wikipedia web pages (105 documents with 799 links).

We begin with an example of the strong relationships between topics and links in the Wikipedia dataset learned by LTHM. The Wikipedia dataset is a collection of 105 web pages with 799 links between the pages in the dataset. We downloaded these web pages from Wikipedia by crawling within the Wikipedia domain, starting from the NIPS[2] Wikipedia page. We have made the data set available online at: http://www.cs.huji.ac.il/~amitg/lthm.html. We used a vocabulary of 2247 words and trained LTHM with 20 hidden aspects. Figure 4 shows four of the hidden topics found by the model in the Wikipedia dataset. For each topic we show the ten most probable words and two most probable links. Topic 1 discusses neural networks, and the two most related links to it (links with high probability to be generated from the topic). Similarly, topic 2 is about speech and pattern recognition. Topic 3 is about cities (Denver and Vancouver, the current and previous venues of the nips conference). topic 4 is about cognitive science and neuroscience. All these topics have related links. Due to lack of space, we show only four example topics, but all 20 topics have clear interpretation and relevant suggested links. A complete list of the topics with top words and top links can be found at http://www.cs.huji.ac.il/~amitg/lthm.html.

We found that the LDA topics on this dataset were of comparable quality, but the assignment of topics to documents can be different in LDA and LTHM, especially for short documents. Table 1 shows a comparison of the document topic vector $\theta_d$ for two short Wikipedia documents, "Journal of Machine Learning" and "Random Forests", in LDA and LTHM. All topics with $\theta_d > 0.05$ are shown. Since LTHM uses the link information, it assigns more weight to the relevant topics.

For quantitative evaluation, we compare LTHM vs. the topic models link-PLSA [6] and link-LDA [9] and a frequency-based method in the task of link prediction. The frequency-based method ranks the documents in the corpus according to the number of time they were linked to from other documents. This ranking serves as the link prediction for all the documents. This prediction is the same for all the documents in the corpus and does not depend on the topics of the source document.

For these experiments we use the Wikipedia dataset and the webkb dataset. The webkb dataset (available online at: http://www.cs.cmu.edu/~webkb) consists of 8282 html pages. For this dataset, we used a dictionary of 2304 words (built according to their frequency in the data and removing stop words). We extracted 12911 links where both ends of the links belong to the webkb corpus. We split each data set into a train set consisting of 90% of the documents and a test set of the remaining 10%. During training, the text of all the documents is provided to all the algorithms, but only the links originating from the documents in the train set are visible during training. Both dataset are learned with 20 hidden aspects. During test, for each test document $d$ we sort all documents $d'$ in the corpus according to the probability of having a link from $d$ (outgoing from any word in $d$) to $d'$.

Figures 3 and 5 show several measures of the performance of the different algorithms. The first measure is the percentage of documents in the test set for which at least one link prediction among the top $N$ is a true link. The motivation for this measure is the following question: Suppose we want to suggest to an author of

---

[2]At the time being, there is no UAI Wikipedia page.

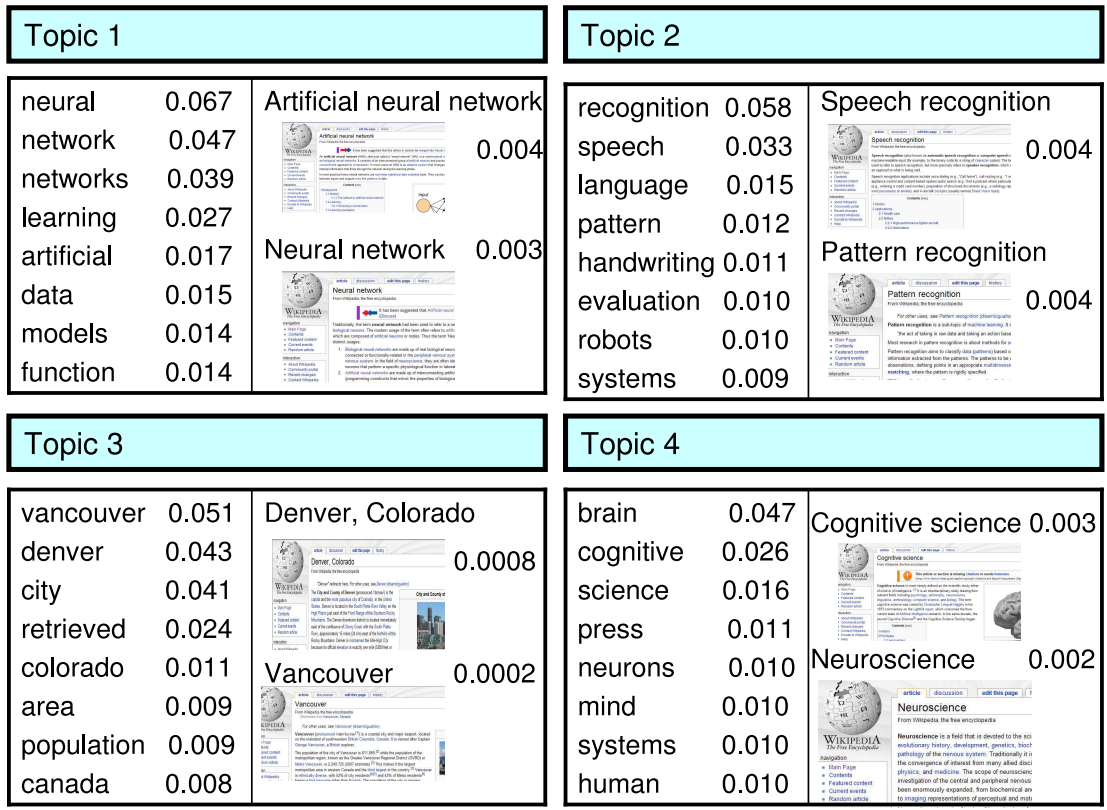

Figure 2: Four example topics learned by LTHM and the links related to them. For each topic, the ten most probable words are shown along with the two links most probable to be generated from that topic. Next to each word and each link is its probability to be generated from the given topic.

Journal Of Machine Learning Research.html

| | LDA | | LTHM |
|---|---|---|---|
| topic prob | top words | topic prob | top words |
| 0.1504 | "search","article","navigation" | 0.4136 | "learning","machine","engineering" |
| 0.0798 | "press","university","new" | 0.0943 | "card","conference","credit" |
| 0.0652 | "learning","machine","algorithms" | | |
| 0.0594 | "fixes","skins","import" | | |
| 0.0533 | "model","regression","reasoning" | | |

Random Forests.html

| | LDA | | LTHM |
|---|---|---|---|
| topic prob | top words | topic prob | top words |
| 0.1416 | "probability","distribution","variables" | 0.2076 | "linear","function","training" |
| 0.1194 | "data","mining","predictive" | 0.1921 | "fuzzy","regression", "model" |
| 0.0757 | "learning","machine","algorithms" | 0.1178 | "bayesian", "model", "network" |
| 0.0542 | fixes","skins","import" | 0.0547 | "carlo","monte","genetic" |
| 0.0527 | stock","market","price" | 0.0524 | "learning","machine","engineering" |
| 0.0527 | search","article","navigation" | | |

Table 1: A comparison of the document topic vector $\theta_d$ for two short Wikipedia documents "Journal of Machine Learning" and "Random Forests" in LDA and LTHM. All topics with $\theta_d > 0.05$ are shown. Since LTHM uses the link information, it assigns more weight to the relevant topics.

a web page other documents to link to. If we show this author $N$ suggestions for links, will s/he use at least one of them? The other measures we use are precision (Among the top $N$ predictions, what is the percentage of true links?) and recall (What percentage of the true links are included in the top $N$ predictions?).

Figure 3 shows that LTHM outperforms all three other methods with respect to all three performance measures. Both link-PLSA and link-LDA do worse than the frequency-based method. This result may seem surprising at first, as these methods are more general than the frequency-based method. In particular, they could fit the relative frequency of each document as its probability to be drawn from any topic. In this case, they would predict the same as the frequency-based method. When we inspect the performance of these methods on the train set (figure 4), we see link-PLSA and link-LDA fit better than the frequency-based method. This suggests that link-PLSA and link-LDA overfit due to the large number of free parameters ($KD$) these models have for modeling links. LTHM, on the other hand, has only $D$ additional parameters for modeling links. Moreover, link generation probabilities depend on the topic mixtures of the documents at both ends of the link. Unlike link-PLSA and link-LDA, no values of the link parameters $\lambda$ can cancel this dependency.

Figure 5 shows the performance of the four methods on the webkb test set. Once again, LTHM outperforms the other methods. The frequency-based method outperforms link-PLSA and link-LDA.

As mentioned in section 3, thanks to the symmetries in LTHM, each EM iteration is computed in time linear in the size of copus times the number of topics. Training on the webkb dataset with 20 topics took 17 hours for 600 EM iterations. Training on the smaller Wikipedia dataset with 20 topcis took 30 minutes for 300 EM iterations.

## 5  Discussion

In this work we have presented LTHM, a novel topic model for hypertext documents. In LTHM, the generation of hyperlinks depends on the topics of the source word and the target document of the link, as well as the relative importance of the target document. Compared to previous approaches, LTHM introduces a much smaller number of additional link parameters. As a result, LTHM achieves good generalization results in cases where other models overfit and fail to generalize.

## Acknowledgements

Support from the Israeli Science Foundation is gratefully acknowledged.

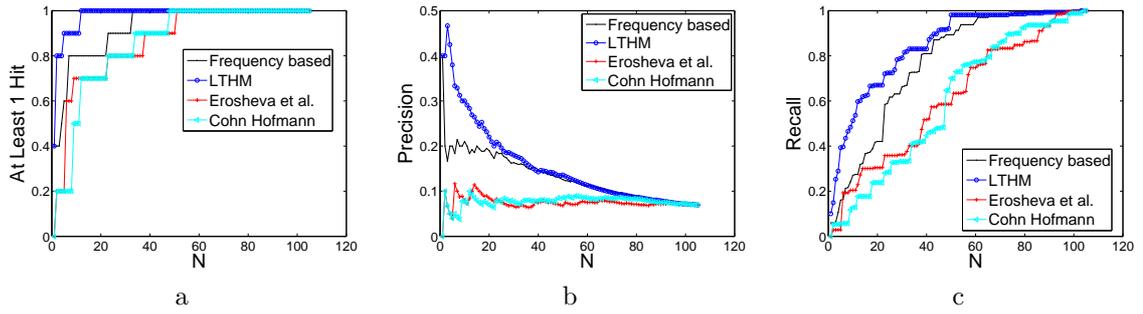

Figure 3: Comparison of link prediction in the Wikipedia test set between LTHM, link-PLSA [6], link-LDA [9] and a frequency-based method. **a.** The percentage of text documents for which there is at least one true link among the first $N$ predicted links. **b.** Average precision for the three methods. **c.** Average recall. LTHM outperforms the other methods, while link-PLSA and link-LDA do worse than the frequency-based method, possibly due to overfitting. See figure 4 and details in the text.

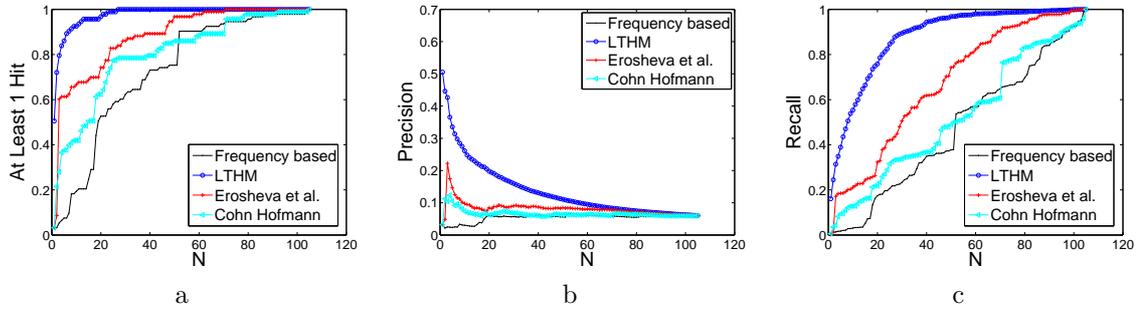

Figure 4: Link prediction in the Wikipedia train set as a measure of parameter fitting by the different methods. Link-PLSA and link-LDA outperform the frequency-based method on the train set, but do worse on the test (figure 3). This suggests overfitting of these methods. **a.** The percentage of text documents for which there is at least one true link among the first $N$ predicted links. **b.** Average precision for the three methods. **c.** Average recall.

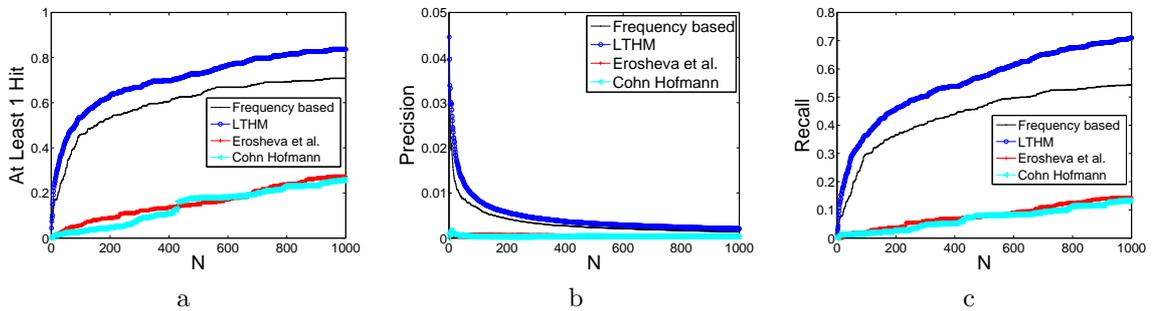

Figure 5: Comparison of link prediction in the webkb test set between LTHM, link-PLSA, link-LDA and a frequency-based method. **a.** The percentage of text documents for which there is at least one true link among the first $N$ predicted links. **b.** Average precision for the three methods. **c.** Average recall. LTHM outperforms the other methods.

## Appendix: Efficient Computation of $E(U_{d,z})$

Naïve computation of $E(U_{d,z})$ requires the computation of $D$ posterior probabilities for each word in the corpus. Taking advantage of symmetries in the model, the aggregation $E(U_{d,z})$ can be computed in $O(K \sum_d N_d)$ (corpus size × number of topics):

$U_{d,z}$ is the expected number of times that $\tau_{d',i} = d$ for any $d', i$ where $z^L_{d',i,d} = z$ but no link has been created (because $z^W_{d',i} \neq z$). Let $P$ denote the parameters of the model, let $O$ denote the full set of observations, and let $\hat{O}$ denote the full set of observation except for the existence or non-existence of the link under discussion. By definition,

$$E(U_{d,z}) \qquad (10)$$
$$= \sum_{(d',i)\in B} \sum_{z'\neq z} \Pr(\tau_{d',i}=d, z^L_{d',i,d}=z, z^W_{d',i}=z'|O,P)$$

where $B = \{(d',i) : no-link(d',i)\}$. It can be shown that $E(U_{d,z})$ can be written as

$$E(U_{d,z}) = \qquad (11)$$
$$\lambda_d \theta_d(z) \left[ \sum_{(d',i)\in B} \frac{(1-\Pr(z^W_{d',i}=z|\hat{O},P))}{\Pr(no-link(d',i)|\hat{O},P)} \right]$$

The probability $\Pr(no-link(d,i)|\hat{O},P)$ depends on the distribution $\Pr(z^W_{d,i}=z|\hat{O},P)$. The latter one can be easily computed from the previously computed posterior distribution of topics of words given all the observations.

$$\Pr(no-link(d',i)|\hat{O},P) \qquad (12)$$
$$= 1 - \sum_z [\sum_d \lambda_d \theta_d(z)][\Pr(z^W_{d',i}=z|\hat{O},P)]$$

To efficiently compute all the expectations $E(U_{d,z})$, one has to follow these steps:

1. Compute $\sum_d \lambda_d \theta_d(z)$ for all $z$ in O(DK).

2. Compute $\Pr(z^W_{d',i}=z|\hat{O},P)$ from $\Pr(z^W_{d',i}=z|O,P)$, then compute $\Pr(no-link(d',i)|\hat{O},P)$ for all $d',i$ in $O(K \sum_d N_d)$ (the number of topics times the size of the corpus).

3. Compute the inner brackets in equation 11 for all topics $z$ in $O(K \sum_d N_d)$.

4. Compute $E(U_{d.z})$ for all $d, z$ according to equation 11 in $O(KD)$.